\documentclass[12pt,preprint]{aastex}

\shorttitle{AGN clustering in the local Universe}
\shortauthors{Cappelluti et al.}

\begin{document}


\title{AGN clustering in the local Universe: an unbiased picture from   {\em Swift}-BAT.}

\author{N. Cappelluti\altaffilmark{1}, M. Ajello\altaffilmark{2,3}, D. Burlon\altaffilmark{1}, M. Krumpe\altaffilmark{4},  T. Miyaji\altaffilmark{5,7}, S. Bonoli\altaffilmark{6}, J. Greiner \altaffilmark{1}}

\altaffiltext{1}{Max Planck Institut f\"ur Extraterrestrische Physik,~D-85478 Garching, Germany}
\altaffiltext{2}{SLAC National Accelerator Laboratory, 2575 Sand Hill Road, Menlo Park, CA 94025, USA}
\altaffiltext{3}{KIPAC, 2575 Sand Hill Road, Menlo Park, CA 94025, USA}
\altaffiltext{4}{  University of California, Center for Astrophysics and Space Sciences, San Diego, 9500 Gilman Dr., La Jolla, CA 92093, USA}
\altaffiltext{5}{Instituto de Astronom\'ia, Universidad Nacional Aut\'onoma de M\'exico,
Ensenada, M\'exico (PO Box 439027, San Ysidro, CA 92143, USA). }
\altaffiltext{6}{Max Planck Institut f\"ur Astrophysik,~D-85478 Garching, Germany}
\altaffiltext{7}{Visiting scholar at University of California San Diego} 
\begin{abstract}
We present the    clustering measurement of hard X-ray selected AGN 
in the local Universe. We used a 
sample of  199 sources  spectroscopically confirmed detected by {\em Swift}-BAT in its 15-55 keV all-sky survey. 
We measured the real space projected auto-correlation function and detected a signal
significant on projected scales lower than 200 Mpc/h. We measured a correlation 
length of r$_0$=5.56$^{+0.49}_{-0.43}$ Mpc/h and a slope $\gamma$=1.64$_{-0.07}^{-0.08}$. 
We also measured the auto-correlation function of Type I and Type II AGN and found 
 higher correlation length for  Type I AGN. We have a marginal  evidence of 
luminosity dependent clustering of AGN, as we detected a larger correlation length 
of luminous  AGN than that of low luminosity sources. The corresponding typical host  DM 
halo masses  of {\em Swift}-BAT are $\sim$ log(M$_{DMH}$)$\sim$ 12-14 h$^{-1}$M/M$_{\odot}$, depending 
on the subsample. For the whole sample we measured    log(M$_{DMH}$)$\sim$ 13.15 h$^{-1}$M/M$_{\odot}$
which is the typical mass of a galaxy group.
We estimated that the local AGN population has a typical lifetime $\tau_{AGN}\sim$0.7 Gyr, it is powered  by 
SMBH with mass  M$_{BH}\sim$1-10$\times$10$^{8}$ M$_{\odot}$  and accreting with very low 
efficiency,  log($\epsilon$)$\sim$-2.0. We also conclude that local
AGN host galaxies are typically red-massive  galaxies with stellar mass of the order 2-80$\times$10$^{10}$ h$^{-1}$M$_{\odot}$. 
We  compared our  results with clustering predictions of  merger-driven AGN triggering   models and found
a  good agreement. 
\end{abstract}


\section{Introduction}
It is now  commonly believed that almost all galaxies host a central 
supermassive Black Hole (SMBH). Dynamical evidence show that the mass of 
the central BHs are closely linked to the mass as well as the 
stellar velocity dispersion of the bulge component of the host galaxy 
(Kormendy \& Richstone 1995, Magorrian et al. 1998).
 This  suggests that the formation and evolution of the 
spheroidal component of galaxies and their SMBH are closely connected.
It is of  upmost importance to understand the mechanism of 
funneling  interstellar gas into the  vicinities of the SMBH,
triggering the accretion. Galaxy mergers or
tidal interaction between close pairs may have played a major 
role (Hopkins et al. 2007), furthermore some mechanism internal to the galaxy, like 
galaxy disk instability may be important. Clustering properties
of AGN in various redshifts give an important clue to 
understand which of these mechanisms trigger AGN activities in what 
stage of the evolution of the universe, through, e.g., the mass
of the Dark Matter Halos (DMH) in which they reside which is linked to BH mass 
life time and Eddington rate. Measurements of the AGN clustering 
show that  AGN are typically
hosted by DMH with a mass of 
the order of  log(M)$\sim$12.0-13.5 M/M$_\odot$ 
 (Yang et al. 2006;  Miyaji et al. 2007; Gilli et al. 2009; Coil et al. 2009;
Hickox et al. 2009; Krumpe et al. 2010).
However, these measurements have 
been produced by using  AGN samples obtained by optical and soft X-ray  (i.e. 0.5-10 keV) surveys.
Optical and soft X-ray selections miss a major part of the SMBH accretion.    
In the optical band the selection of AGN is biased by
galaxy starlight dilution and by dust absorption. 
Although luminous soft X-ray emission is  a signature  of the presence of an 
AGN,  absorbed sources can  be missed with a soft X-ray selection, either because
 they are intrinsically less luminous 
 (Hasinger et al. 2008) or
because of the high column density. However, X-ray emission from these
sources leaks out at higher energies (i.e $>$ 5-10 keV)
where the efficiency  of instruments mounting X-ray focusing optics is 
low. For this reason hard X-ray  selected samples could provide clean and 
unbiased samples of AGN.  
The {\em Swift}-BAT 
all-sky survey provides a spectroscopically complete (100 \%) sample of local AGN detected 
 in the 15-55 keV energy  band, 
with an unprecedented depth and characterization of the source properties,
from redshifts to column densities.
In this letter we present  the first   
 study of  clustering of hard X-ray selected  AGN in the local Universe. 
Throughout this paper we will assume a $\Lambda$-CDM cosmology with 
$\Omega_m$=0.3, $\Omega_{\Lambda}$=0.7, H$_0$=100$h^{-1}$ km s$^{-1}$ Mpc and 
$\sigma_8$=0.8. Unless otherwise stated errors are quoted at the 1$\sigma$ level.

\section{The Sample of Swift BAT hard X-ray selected AGN }

The Burst Alert Telescope 
(BAT; Barthelmy et al. 2005)
on board the {\em Swift} satellite 
(Gehrels et al. 2004), represents
a major improvement in sensitivity for imaging the hard X-ray sky.
BAT is a coded mask telescope with a wide field of view
(FOV, 120$^\circ\times$90$^{\circ}$ partially coded) aperture sensitive in
the 15--200\,keV domain. Thanks
to its wide FOV and its pointing strategy, BAT monitors continuously
up to 80\% of the sky every day achieving, after several years of 
survey, deep exposure in the entire sky.
Results of the BAT survey 
(Markwardt et al. 2005, Ajello et al. 2008, Tueller et al. 2009)
show that BAT reaches a sensitivity of $\sim$1\,mCrab\footnotemark{}
\footnotetext{1\,mCrab in the 15--55\, keV band corresponds to
1.27$\times10^{-11}$\,erg cm$^{-2}$ s$^{-1}$}
in 1\,Ms of exposure.

The sample used in this work consists of 199 non-blazar AGN detected
by BAT during the first three years and precisely between March 2005
 and March 2008. This sample is part of the one used in Ajello et al. (2009) 
which comprises all sources detected by BAT at  high ($|$b$|$$>$15$^{\circ}$)
Galactic latitude and with a signal-to-noise ratio (S/N) exceeding 5\,$\sigma$.
The reader is referred to Ajello et al. (2009) 
for more details about the sample
and the detection procedure. The flux limit at each direction in the sky has been
computed, following Ajello et al. (2008), 
analyzing the local background around that position.
For each source we use the redshift already provided in Ajello et al. (2009) 
and the measurement of the absorbing column density as determined from joint XMM-Newton/XRT--BAT
fits (Burlon et al, in preparation). The redshift distribution of the sample is shown in Fig. \ref{fig:sample} together with the 
redshift cone diagram of the survey up to 20000 km/s (z$\sim$0.07).

\section{The two-point spatial  auto-correlation function}

The two-point auto-correlation function ($\xi(r)$, ACF)
describes the excess probability over random of finding a pair 
with an object in the volume $dV_1$ and another in the volume $dV_2$, separated by a distance $r$ so that
$dP=n^2[1+\xi(r)]dV_1 dV_2$, where n is the mean space density. A known effect when measuring pairs separations is 
that the peculiar velocities combined with the Hubble flow
may cause a biased estimate of the distance
when using the spectroscopic redshift.  To avoid this effect
we computed the projected ACF (Davies \& Peebles 1983):  
 $w(r_p)=2\int_{0}^{\pi_{max}}\xi(r_p,\pi)d\pi$.
Where $r_p$ is the distance component perpendicular to the line of sight and 
$\pi$ parallel to the line of sight (Fisher et al. 1994).
 It can be demonstrated that, if the ACF is expressed as 
$\xi(r)=(r/r_0)^{-\gamma}$, then
\begin{equation}
  w(r_{\rm p})=A(\gamma)r_{0}^{\gamma}r_{p}^{1-\gamma},
\label{eq:chi}
\end{equation}
 where  $A(\gamma)=\Gamma(1/2)\Gamma[(\gamma-1)/2]/\Gamma(\gamma/2)$ (Peebles 1980).\\
The ACF has been estimated by using the minimum variance estimator 
described by Landy \& Szalay (1993):
\begin{equation}
 \xi(r_p,\pi)=\frac{DD-2DR+RR}{RR}
\label{eq:landy}
\end{equation}
where DD, DR and RR are the normalized  number of
data-data, data-random, and random-random source pairs, respectively. 
Equation \ref{eq:landy}  indicates that an accurate estimate of the 
distribution function of the random  samples is crucial in order to 
obtain a reliable  estimate of $\xi(r_p,\pi)$. 
Several observational biases must be taken 
 into account when generating a random sample of objects in a flux limited survey. 
In particular, in order to reproduce the selection function of the survey,
one has to carefully reproduce the space and  flux distributions 
of the sources, since the sensitivity of the survey in not homogeneous on the sky. 
Simulated AGN were  randomly placed on the  survey area.   
In order to reproduce the flux distribution of the real sample, we followed 
the method described in Mullis et~al.~(2004).  
The cumulative AGN 
logN-logS source count distribution, in the 
15-55 keV band, can be  described by a power law, $S=kS^{-\alpha}$, with 
$\alpha\sim1.55$ (Ajello et al. 2008) 
Therefore, the differential probability scales as $S^{-(\alpha+1)}$. 
Using a transformation method  the 
random flux above  a certain X-ray flux $S_{lim}$ is distributed as 
$S=S_{lim}(1-p)^{\frac{-1}{\alpha}}$, where $p$ is a random number uniformly 
distributed between 0 and 1 and $S_{lim}$=7.6$\times10^{-12}$ erg cm$^{-2}$ 
s$^{-1}$. All  random AGN 
with a flux lower than the flux limit map  at  the source position were  excluded.
Redshift were randomly drawn from the smoothing of the real redshift distribution
with a gaussian kernel with $\sigma_z$=0.3.
 An  important choice for obtaining a reliable estimate of $w(r_{\rm p})$,
is to set $\pi_{max}$ in the calculation of the integral above. 
One should avoid values of  $\pi_{max}$ too large since they would  add noise to
the estimate of   $w(r_{\rm p})$.  If, instead,   $\pi_{max}$ is too small  one could not recover
all the signal. 
We have calculated $w(r_{\rm p})$  by varying   $\pi_{max}$ and
 found that the result converges at $\pi_{max}\sim$60 Mpc/h.\\
Errors on $w(r_{\rm p})$ were computed with a bootstrap resampling technique with 
100 realizations.
 It is worth noting that in the literature, several methods  
are adopted for errors estimates in two-point statistics, and no one has been proved to be the most 
precise. However, it is  known  that Poisson estimators generally underestimate the variance because 
they do consider that points in ACF are not  statistically independent. 
Jackknife resmpling method, where one divides the survey area in many
sub fields and iteratively re-computes correlation functions by
excluding one sub-field at a time, generally gives a good estimates of 
errors. But it requires that sufficient number of almost statistically
independent sub-fields. This is not the case for our small sample.
For these reasons  we used the bootstrap resampling for the error estimate which,
in our case, are comparable with the Poisson errors.

\section{Results}

We show in Fig. \ref{fig:acf} the projected ACF measured on the  whole AGN sample
 of the survey. The ACF has been evaluated in the projected separation range 
$\sim$0.2 Mpc/h $<r_p<$ 200 Mpc/h and has been plotted in form of $w(r_{\rm p})/r_p$ in order to reproduce
the slope of $\xi(r)$ (see above). The bin size for computing $w(r_{\rm p})$ has been set to $\Delta  log(r_p,\pi)$=0.15.
 We obtained an estimate of   $w(r_{\rm p})/r_p$ 
with a significance of the order  4$\sigma$-5$\sigma$.  In order to evaluate the power of 
the clustering signal we fitted  $w(r_{\rm p})$ with $\chi^2$ minimization technique
by using  Eq.\ref{eq:chi}  as a model with  $r_0$ and $\gamma$ as free parameters. 
The correction due to the integral constraint (Peebles 1980)
is estimated to be much smaller than the statistical uncertainties in 
our sample, and thus has not been made.
As a result we obtained $r_0$=5.56$^{+0.49}_{-0.43}$ Mpc/h and $\gamma$=1.64$^{+0.07}_{0.08}$. The confidence
contours of the fit are presented in Fig. \ref{fig:acf}. 
We also  measured  the  ACF for different data subsamples. We first divided the 
sample according to the column density: we defined as Type II AGN (or absorbed) sources with 
  log$(N_{H})\geq$22 cm$^{-2}$   and  as Type I AGN (or unabsorbed) sources if   log$(N_{H})<$22 cm$^{-2}$.
As a result we constructed a sample of 96 Type
 I AGN  and one of 103 Type II AGN. 
For both  samples we computed the ACF with the technique described above. 
We also split the sample into  high and low luminosity
subsamples. All the sources with  L$_{15-55}>$43.2 erg/s (i.e. the median luminosity of  the whole
sample, HL sources) were classified as high luminous,  while the sources with L$_{15-55}<$43.2 erg/s  (LL sources)
as low luminosity sources.  
The results of the measurement of the ACF as a function of the source type and luminosity class are presented in
Fig. \ref{fig:acf} together with the fit confidence contours. 
{Note  that for the HL sample the fit parameters are poorly constrained because
 of the lack of close pairs in the sample.
  We also repeated the fit by freezing $\gamma$ to 1.7, and obtained consistent results (Table \ref{tbl-1}).
A summary of the  fit results  of all the samples used here is given in Table \ref{tbl-1}.
Type I  AGN show a larger correlation 
  with respect to that of type II  AGN, the significance of this difference is of the order $\sim2.7\sigma-3.3\sigma$.
  HL AGN show a 1.7-4.6$\sigma$ higher correlation
length with respect to LL AGN. We also checked the correlation between r$_0$ and $<L_X>$  of all the subsamples and found 
a linear correlation coefficient R=0.95, which corresponds to a $\sim2\sigma$ significant correlation. 
 We can use the weighted mean dispersion of the results on the measurement of r$_{0}$ in our subsamples to 
 estimate the impact of sample variance on our results 
 under the assumption that this is the main reason of the 
 fluctuations. Our estimates suggest that overall our results 
 are affected by this effect at $\sim$10\% level.  
It is worth to note that our results are more significant than those obtained by e.g. Mullis et al. (2004),
with a similar number of sources. This is because our sources 
are distributed in a much smaller volume than that sampled by the NEP survey
and, by being on average less luminous, have an intrinsic higher space density
resulting in a larger number of close source pairs.  \\
\begin{table}[!b]
\tiny
\begin{center}
\caption{Summary of the results.\label{tbl-1}}
\begin{tabular}{ccccccc}
\tableline\tableline
Sample & N$^{a}$ & $<z>$&$< log(L_X)>$ &r$_{0}$ & $\gamma$ & r$_{0_{\gamma=1.7}}^{b}$ \\
& &       & erg/s   & Mpc/h &   &   Mpc/h   \\
\tableline
All    &  199& 0.045 & 43.2 &  5.56$^{+0.49}_{-0.43}$ &  1.64$_{-0.08}^{+0.07}$ & 5.54$^{+0.1}_{-0.1}$\\
Type I & 96& 0.046 & 43.37& 7.93$^{+1.14}_{-0.79}$&2.1$^{+0.20}_{-0.25}$ & 8.12$_{-1.00}^{+1.57}$\\
Type II & 103 & 0.024 & 42.87& 4.72$^{+0.60}_{-0.70}$ &1.78$^{+0.24}_{-0.17}$ & 4.90$^{-0.70}_{+0.20}$\\
HL &99&0.054 &43.67&  13.92$^{+5.48}_{-6.69}$ &1.41$^{+0.15}_{-0.19}$  & 15.63$_{-2.57}^{+1.57}$\\
LL &100&0.023 &42.55&  3.37$^{+0.51}_{-0.68}$&1.86$^{+0.25}_{-0.17}$& 3.56$_{-0.66}^{+0.15}$\\
\tableline\tableline
\end{tabular}
\tablenotetext{a}{Number of sources in the Sample.}
\tablenotetext{b}{r$_{0}$ obtained by freezing $\gamma$=1.7 in the fit.}

\end{center}
\end{table}

\begin{table*}[!t]
\begin{center}
\tiny
\hspace{-.5cm}
\caption{Bias factor and Mass of the Dark matter halos hosts of the AGN in
the samples.\label{tbl-2}}
\begin{tabular}{cccccccccc}
\tableline\tableline
Sample & b$_X^{a}$& M$_{DM}^{b}$&  $\tau_{AGN}^{c}$ &  log(M$_{BH})^{d}$) &  log($\epsilon$)$^{e}$ & M$^{* f}$\\
       &         &                      log(M/ h$^{-1}$M$_{\odot}$) & Gyr   &  log(M/M$_{\odot}$)  &  & 10$^{10}$/M$_{\odot}$\\
\tableline
All     &$1.21^{+0.06}_{-0.07}$ &$13.15^{+0.09}_{-0.13}$  & 0.68  & 8.51 & -1.96 & 18.2\\
Type I  &$2.01^{+0.15}_{-0.13}$ &$13.94^{+0.15}_{-0.21}$  &4.99 & 8.79 & -2.02 & 31.6\\
Type II &  $1.08^{+0.26}_{-0.29}$ &$12.92^{+0.11}_{-0.38}$  &1.32 & 7.96 &  -1.85 & 6.38\\
HL      &   $2.28^{+0.95}_{-0.90}$ &$14.08^{+0.37}_{-0.70}$ & 3.91 & 9.28 & -2.12 & 80.5 \\
LL      &      $0.80^{+0.06}_{-0.16}$ &$11.89^{+0.34}_{-\infty}$ &  0.24&   7.43 & -1.68 & 2.28  \\
\tableline
\end{tabular}
\tablenotetext{a}{AGN bias factor}
\tablenotetext{b}{Mass of the typical Dark matter halo hosting an AGN}
\tablenotetext{c}{AGN duty ciycle in Gyr}
\tablenotetext{d}{Black hole mass}
\tablenotetext{e}{Eddington ratio }
\tablenotetext{f}{Stellar mass of the bulge }
\end{center}
\end{table*}
In the linear theory of  structure formation, the bias factor
defines the  relation between the autocorrelation function of large scale structure
tracers and the underlying overall matter distribution. In the case of X-ray selected 
AGN, we can define the following relation: $\xi_{X}(r,z)=b_{X}(r,z)^2\xi_m(r,z)$, where $\xi_X$,
$\xi_m$ and b$_X$ are the autocorrelation function of AGN, of DM and the AGN bias 
factor, respectively.
In order to compute the b$_X$, we estimated  the amplitude of the 
fluctuations of the AGN distribution in a sphere of 8 Mpc/h (also know as $\sigma_8$),
by using Eq. 12 and 13 of Miyaji et al. (2007) 
and  all the combinations of  r$_{0}$ and 
$\gamma$ are reported in Tab. \ref{tbl-1}. In order to derive the bias factor of the AGN 
in our samples  we used $b_{AGN}=\sigma_{\rm 8,AGN}(z)/\sigma_8 D(z)$, 
where D(z) is the growth factor. This quantity allows us to compare the
observed AGN clustering to the underlying mass distribution from linear
growth theory (Hamilton 2001). 
As a result we obtain for the whole sample 
b$_{AGN}$(z$\sim$0.04)=1.21$^{+0.06}_{-0.07}$. 
We have repeated this calculation for all the samples listed in Table \ref{tbl-1} 
and we report the corresponding values of the bias factor in Table \ref{tbl-2},
for all the possible best fit results. \\
It is widely accepted that the clustering amplitude of DMH
depends primarily on their mass (see e.g. Sheth et al. 2001). 
In this way, we can estimate the typical mass of the DMH in 
which the population of AGN reside, 
under the assumption that the typical mass of the host 
halo is the only variable that  causes AGN biasing.
We have then computed the expected large-scale
bias factor for different dark matter halo masses by using the 
prescription of  Tinker  et al. (2005). 
The required ratio of
the critical overdensity to the rms
fluctuation on a given size and mass is calculated by $\nu = \delta_{\rm
cr}/\sigma(M,z)$, for our purposes we assumed $\delta_{\rm cr}\approx 1.69$ and
compute $\sigma_8(M,z)$ and therefore $b(M,z)$ using Eq.~A8, A9, and A10 
in van den Bosch (2002). 
The typical DMH mass that hosts an AGN 
 has been estimated to be
$ log(M_{DMH})$=13.15$^{+0.09}_{-0.13}$ h$^{-1}$M/M$_{\odot}$.
This is consistent 
 with similar measurements in the local Universe  of Krumpe et al. (2010), Grazian et al. (2004) and Akilas et al. (2000).
We have computed the typical mass of the DMH  for all the subsamples 
listed in Table \ref{tbl-1} and reported for simplicity in Table \ref{tbl-2}.
 Following Martini \& Weinberg(2001), by knowing the AGN and DMH
halo space density at a given luminosity and mass ($n_{AGN}, n_{DMH}$), one can estimate
the duty cycle of the AGN, $\tau_{AGN}(z)=\frac{n_{AGN}(L,z)}{n_{DMH}(M,z)}\tau_H(z)$. Where $\tau_H(z)$
is the Hubble time at a given redshift\footnote{ This is an upper limit obtained by assuming that the lifetime of the 
DM halo is of the order of $\tau_H(z)$ }. For the whole sample at z$\sim$0,
  $n_{DMH}\sim$6.7$\times$10$^{-4}$ Mpc$^{-3}$  (Hamana et al. 2002) and 
 $n_{AGN}\sim$3.4$\times$10$^{-5}$ Mpc$^{-3}$  (Sazonov et al. 2007) which leads to an
estimate of $\tau_{AGN}(z=0)\sim$0.68 Gyr.  To fully characterize our sample, we derived 
the average properties of the active BHs and their host galaxies.
By using the bolometric correction prescribed by Hopkins et al. (2007) we estimated from 
$<L_X>$, $<L_{Bol}>$ and $<L_B>$ (B band luminosity).  L$_B$ is related to the 
black holes mass and the stellar mass of the host galaxy via scaling relations (Marconi \& Hunt 2003).  
From $<M_{BH}>$ we derived $<L_{Edd}>$ and the Eddington rate $\epsilon$=$<L_{Bol}>/<L_{Edd}>$ (see Table \ref{tbl-2} for a summary).
 We  point out that the 
estimates obtained above have several limitations which mostly arise from the uncertainties on
scaling relations and from the broad range of luminosities sampled here. We therefore consider
these values as  estimate for the ``average'' local AGN population.   
\section{Summary and discussion} 
In this letter we report on the measurement of clustering of 199 AGN in the local Universe using
 the Swift/BAT all-sky survey sample. 
This result gives, for the first time, an unbiased picture of the 
z=0 DMH-galaxy-AGN coexistence/evolution. 
We  obtained  a correlation length r$_0$=5.56$^{+0.49}_{-0.43}$  Mpc/h  and
$\gamma$=1.64$_{-0.08}^{+0.07}$. We  measured the ACF 
for Type I and Type II AGN and found a significant difference in their
correlation lengths.  We have measured a marginally significant
higher r$_0$ for high luminosity AGN than for the low luminosity ones.
We propose that the observed difference in Type I vs. Type II clustering is driven 
by the intrinsic higher $<L_X>$ of Type I AGN as we show a marginal evidence
of a correlation between r$_0$ and $L_X$.
We estimated the typical mass of the DMH hosting an AGN
of the order  log(M$_{MDH}$)$\sim$13.15 h$^{-1}$M/M$_{\odot}$. 
 In Fig.  \ref{fig:bias} we show  the  bias-redshift  plane
results from AGN and galaxy surveys (references in the figure).
 In the same plot we show the expected evolution
 of different DMH masses.
We compared only bias values of studies
that rely on the real space correlation function $\xi(r)$
(values of $\sigma_{\rm 8, AGN/GAL}$  from Krumpe et al.  2010). 
This approach allows us to compare all different clustering studies on a common basis.

The majority of the X-ray  surveys agree with a picture where AGN are typically hosted in 
DM halos with mass of the order of 12.5 h$^{-1}$M/M$_{\odot}<$ log(M$_{MDH}$)$<$13.5  h$^{-1}$M/M$_{\odot}$ which is the mass
of moderately poor group. Optically selected AGN instead reside in lower density environment and of the order
of the  log(M$_{MDH}$)$\sim$12.5  h$^{-1}$M/M$_{\odot}$.  
Another interesting aspect
is that   X-ray selected AGN samples (including ours)  cluster similarly to red galaxies and that LL AGN or type II 
AGN are found typically in less massive environments. On the contrary HL AGN and Type I AGN are hosted in  massive 
galaxies in massive DM halos (clusters).
 
We estimated that Swift-BAT AGN are powered by black holes with a typical ~mass  log(M$_{BH}/M_{\odot}$) $\sim$8.5,
accreting at very low Eddington ratio (i.e. $\sim$0.01 L$_{Edd}$) and that they are hosted by massive galaxies with 
mass of the order $\sim2\times10^{11}$ M/M$_{\odot}$. These properties, except $\epsilon$,
 scale with  $<L_X>$ and Type (HL and Type I are hosted in higher mass DMH in  more massive galaxies with bigger black holes).
 The upper limits on the duty cycle suggest that these AGN are shining since at least for 0.2-1.2 Gyr \footnote{HL and type I estimate may be 
 wrong because of the relatively young age of 10$^{14}$  (M/$M_{\odot}$) DMH}.  


We then tested the AGN merger-driven triggering paradigm by comparing the 
theoretical predictions for AGN clustering of the model of  Marulli et al. 
(2008) and Bonoli et al. (2009) with our results on the whole sample. The  
theoretical model is based on the assumption that AGN activity is triggered 
by galaxy mergers and the lightcurve associated to each accretion event is 
described by an Eddington-limited phase followed by a quiescent phase modeled 
after Hopkins et al. (2005).  Applied to the Millennium Simulation (Springel et al. 2005), 
such model has been shown to be successful in 
reproducing the main properties of the black hole and AGN populations 
(Marulli et al. 2008) and the clustering of optical quasars (Bonoli et al. 
2009).  Using this model, we computed the expected correlation length 
of a sample of  simulated  AGN at z$\sim$0 with intrinsic $L_{bol}$ 
luminosities similar to the ones of our observed AGN. The model predicts a 
clustering  length  r$_0$=5.68$\pm{0.08}$
which is in agreement  within 1$\sigma$ with our measurement.
  
   By merging the observational evidences 
 and the model predictions, a plausible scenario   for the 
 history of local AGN is  the following:
 \begin{itemize}
 \item {\em Swift-BAT} AGN switched on about 0.7 Gyr Ago after a galaxy merger event.
 \item  They shine in an Eddington-limited regime for the first part of their lives
 	   where they gain most of their mass.	
\item  In the second phase of their lives (i.e. after 0.2-0.5 Gyr) they start to accrete with lower and lower 
efficiency. Their luminosity drops because of the decreased gas reservoirs. 
\item At z$\sim$0 they have grown to $\sim$10$^{8-9}$ M$/M_{\odot}$ SMBHs,
               shining as moderately low-luminosity  AGN at low accretion rates.  
\end{itemize}
\acknowledgments
NC thanks Gigi Guzzo, Roberto Gilli , Angela Bongiorno and  Simon White for the useful comments.
Support from NASA NNX07AT02G, CONACyT 83564, PAPIIT IN110209 is aknowledged.

\begin{figure}[!t]
\begin{center}
\includegraphics[width=.45\textwidth]{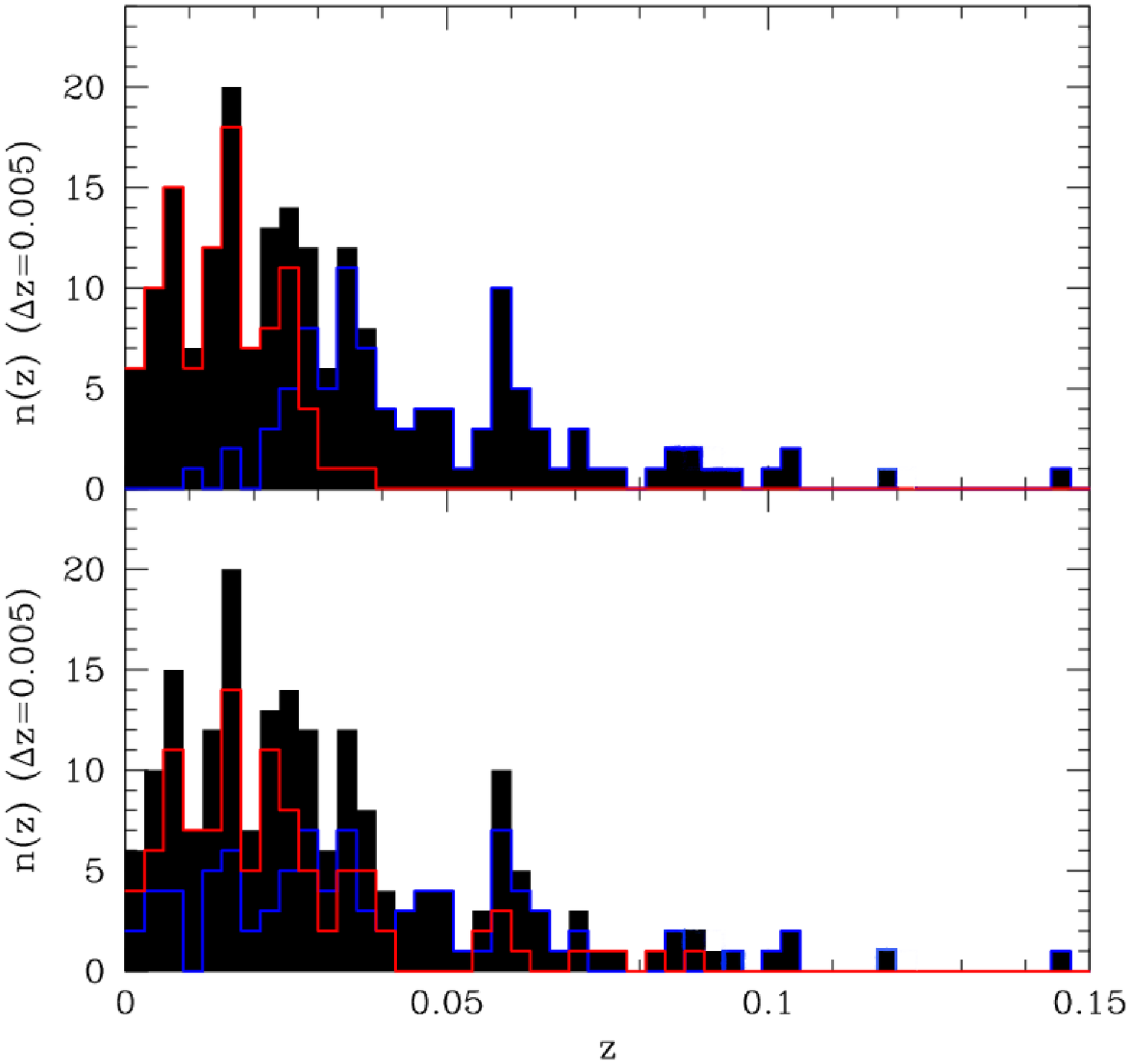}
\includegraphics[width=.45\textwidth]{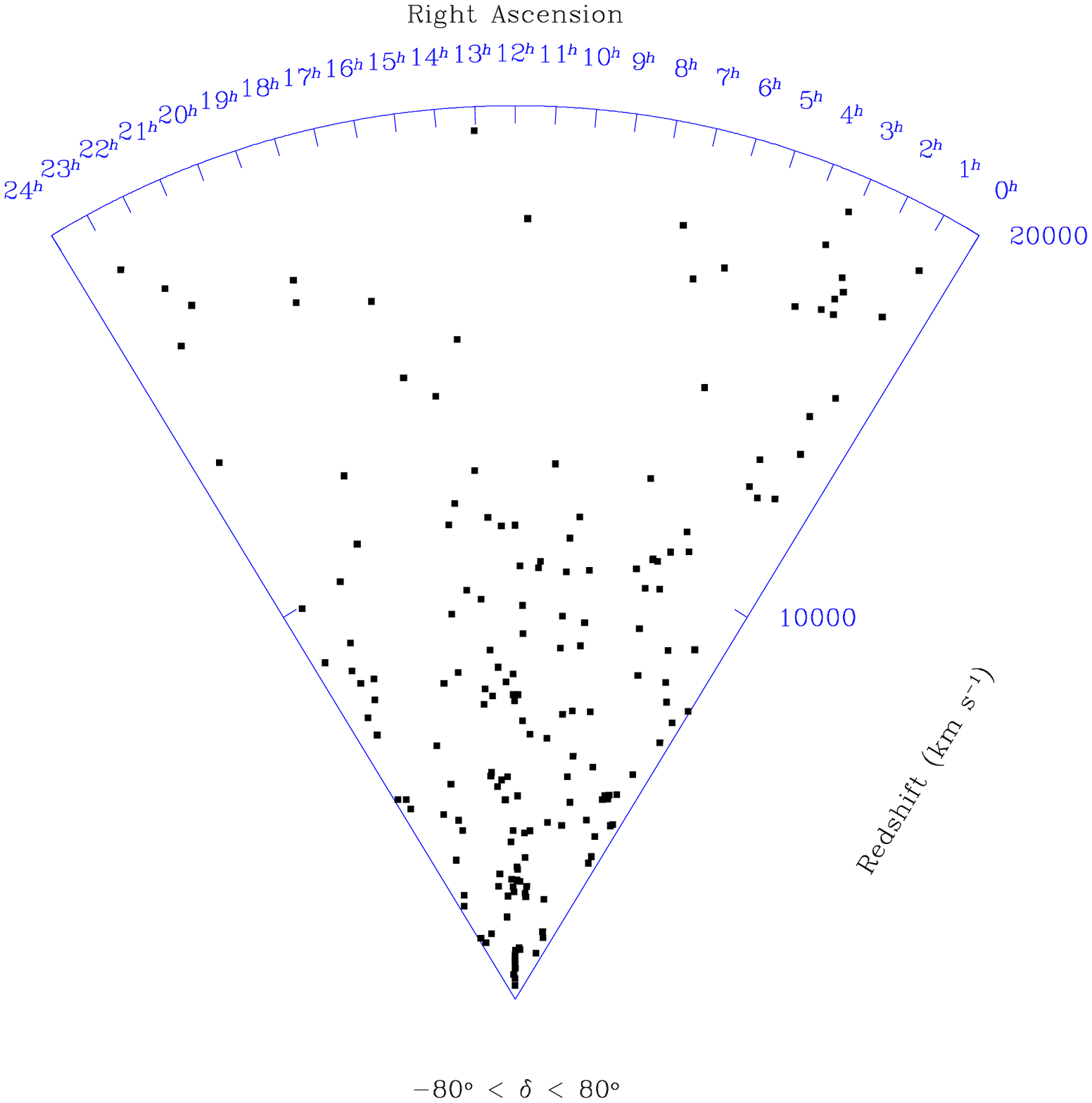}
\caption{$Upper~Left~Panel$: The redshift distribution of HL AGN ($blue$) and LL AGN ($red$) overlaid to the full sample ($black$).    
$Bottom~Left~Panel$: The redshift distribution of Type I AGN ($blue$) and Type II AGN ($red$) overlaid to the full sample ($black$). 
$Right~Panel:$ A representation of the spatial distribution of local AGN with a redshift
cone diagram of the {\em Swift)}-BAT  AGN    in the declination interval -80$<\delta<$80.      \label{fig:sample} }
 
\end{center}
\end{figure}

\begin{figure}[!b]
\begin{center}
\includegraphics[width=\textwidth]{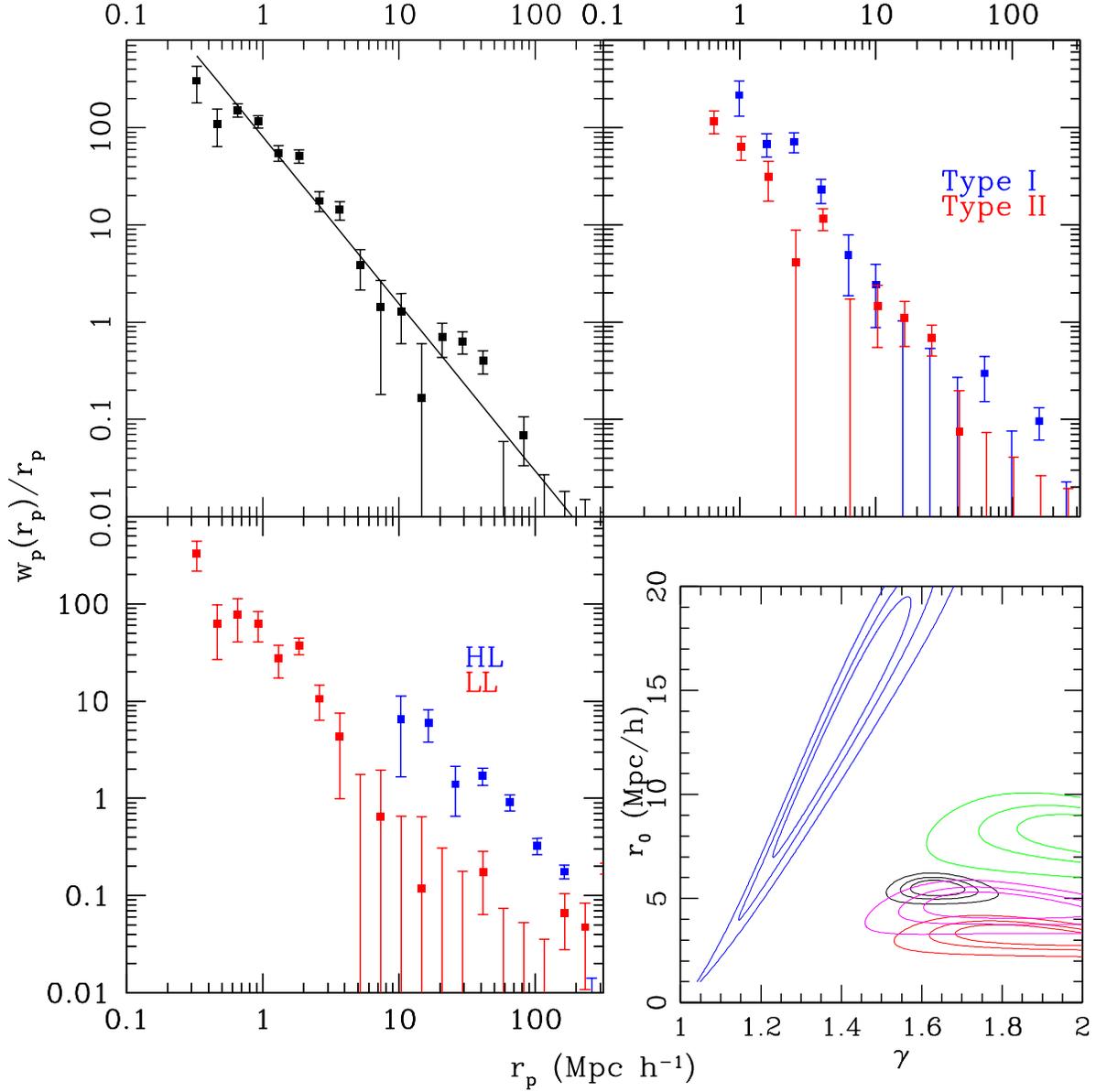}
\caption{$Upper~Left~Panel:
$ $w(r_{\rm p})/r_p$  for the whole sample, the solid line represents the best fit power-law.  
    $Upper~Right~Panel:$  The ACF of type  I AGN ($blue$) and Type II ($red$). $Bottom~Left~Panel:$ The ACF of HL     ($blue$) and LL ($red$) AGN.    $Bottom~Right:$ Confidence contours of the two parameter fit to $\xi(r)$ for the whole sample ($black$), Type I AGN ($green$), type II AGN ($pink$), HL AGN ($blue$) and LL ($red$). \label{fig:acf}}

\end{center}
\end{figure} 

\begin{figure}[!h]
\begin{center}
\includegraphics[width=\textwidth]{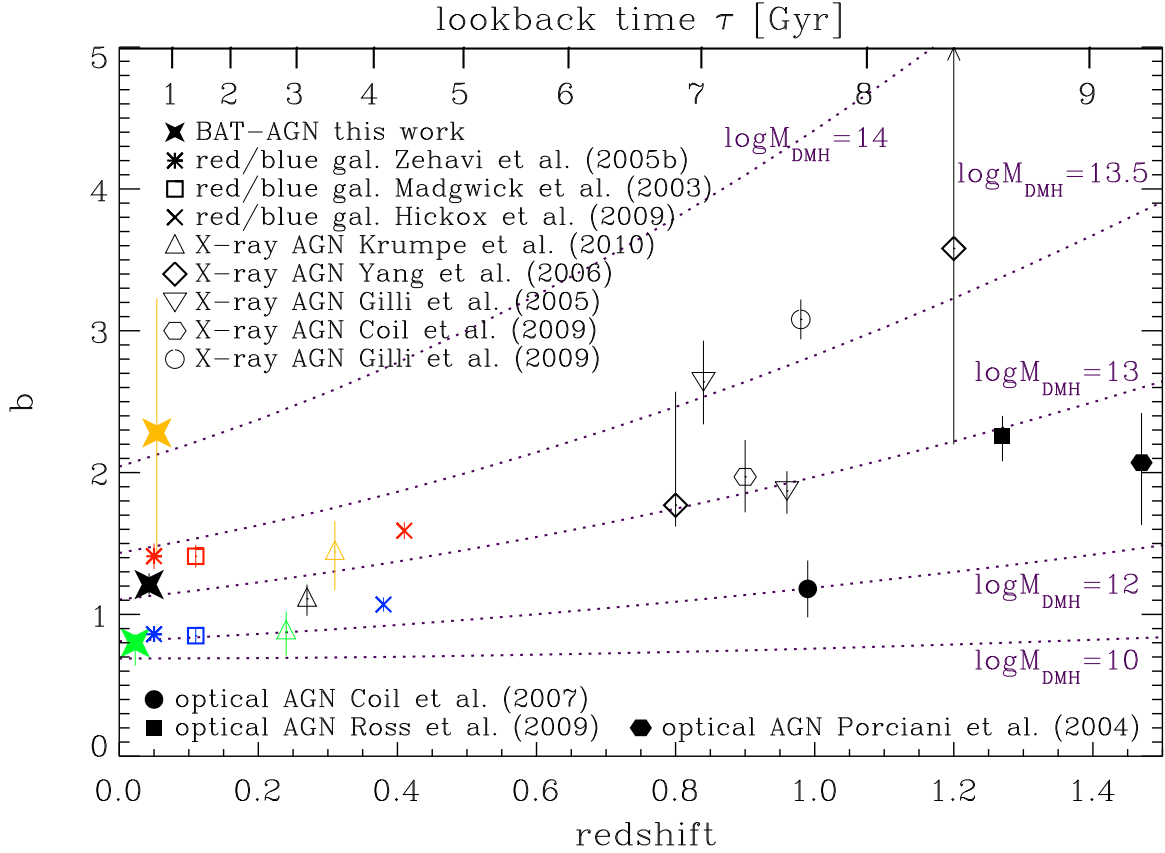}
\caption{Bias parameter $b_{\rm AGN}$ as a 
         function of redshift for studies that are based on the real space 
         correlation function. Yellow (green) symbols represent high (low)luminosity subsamples 
         of the corresponding study. The blue points for galaxy clustering 
         measurements belong to blue galaxies, while red points represent 
         the red galaxy sample. The dotted lines give the expected $b(z)$ of 
         typical dark matter halo masses based on Sheth et al. (2001) 
         and van den Bosch (2002). 
         The masses are given in  log $M_{\rm DMH}$ in units 
         of $h^{-1}$ $M_\odot$.\label{fig:bias}}
\end{center}
\end{figure}

\end{document}